a shortened version appeared
in *Nature* **407**, 55-57 (2000)# Non-Quantized Penetration of Magnetic Field in the Vortex State of Superconductors: "Fractional" and "Negative" Vortices

A.K. Geim[†&], S.V. Dubonos[†*], I.V. Grigorieva[†&], K.S. Novoselov[†], F.M. Peeters[o] & V.A. Schweigert[o+]

[†]University of Nijmegen, Toernooiveld 1, 6525 ED Nijmegen, The Netherlands
[&]Department of Physics, The University of Manchester, M13 9PL Manchester, United Kingdom
[*]Institute for Microelectronics Technology, 142432 Chernogolovka, Russia
[o]Department of Physics, University of Antwerpen (UIA), B-2610 Antwerpen, Belgium
[+]Institute of Theoretical and Applied Mechanics, 630090 Novosibirsk, Russia**Flux quantization is an essential attribute of superconductivity, and it is commonly assumed that the magnetic field enters type-II superconductors as an integer number of vortices, each carrying one flux quantum, $\phi_0 = h/2e$. This assumption is generally wrong. As first pointed out by Bardeen and Ginzburg in the early sixties[1,2], the amount of flux carried by vortices depends on their distance to the sample edge and can be smaller than $\phi_0$. In bulk superconductors, the reduction is negligible already at submicron distances from the edge but, in thin films, the effect may survive much farther[3,4]. However, in the absence of any experimental observation, such flux reduction is perceived to be an exotic or negligible effect. Here we report the first measurements of the amount of flux introduced by individual vortices in a superconducting film, finding that the flux always differs substantially from $\phi_0$. We have observed vortices that carry as little as $0.001\ \phi_0$ as well as seemingly "negative vortices" whose penetration leads to the expulsion of magnetic field. We distinguish two phenomena responsible for non-quantized flux penetration: the finite-size effect[1-4] and a non-linear screening of magnetic field in the presence of a surface barrier. The latter effect has not been considered previously but is likely to cause non-quantized penetration in most cases.**Magnetic properties of superconductors, including their critical currents, are determined by flux motion through a superconductor as a whole, which involves flux propagation not only through the bulk but also through the superconductor's edge. Because of the inevitable pinning in real superconductors, vortices initially penetrate only at a finite (usually, mesoscopic) distance from the edge. This effectively creates an edge layer that serves as a reservoir of vortices that are subsequently injected further into the bulk, and there is growing evidence that such a layer can significantly influence global superconducting properties of a material (e.g., refs. 5,6). On the other hand, vortices near the edge are not exactly the same as vortices in the bulk: they are expected to have a rather different structure and energy because the distribution of their electric currents is modified by the presence of the edge[1-4]. Such changes can be particularly important in the case of thin films, where the effective magnetic penetration length $\lambda_{eff}$ becomes large and screening is non-exponential[3,4].

Among most directly observable consequences of the edge influence on the vortex structure is that vortex's flux is no longer quantized and becomes smaller than $\phi_0$ [1-4]. Although the flux reduction is theoretically known for several decades, such vortices carrying a minor fraction of $\phi_0$ have never been observed or inferred in an experiment. This provided the original motivation to our work, as we found a way to address the issue by making use of the ballistic Hall magnetometry. This technique[7,8] allows magnetization measurements with the resolution much better than one flux quantum on micrometer-sized samples, where the influence of the edge is dominant.

Figure 1 plots a typical behaviour we observe for the initial stages of the field penetration into relatively large thin-film samples (in this case, samples have diameter $D \approx 15\mu m$ and thickness $h \approx 0.1\mu m$). Note that disks of this size adequately represent the corresponding properties of a macroscopic film. Indeed, due to bulk pinning, which is rather weak[8] but still present, the entering vortices jump no farther than a few microns from the sample's edge (we observe hysteresis due to bulk pinning if $D > 4\mu m$). This makes the measurements for the disks in Figure 1 essentially equivalent to a study of flux penetration into a $\pi D$-long strip of the same macroscopic film.

Magnetic flux penetrates inside disk (a) in a sequence of steps so that each of the steps corresponds to a vortex or a number of vortices jumping inside, in agreement with general expectations. However, a more careful look reveals that the step height is not quantized and some of the jumps are smaller than $\phi_0$. We postpone a discussion of this observation and refer directly to another (nominally similar) sample in Fig. 1. Disk (b) reveals a completely different picture that, however, has been observed for many samples. Here, after the initial region of the full Meissner effect, the flux enters inside the film relatively smoothly and only after several (five or so) flux jumps, the behaviour becomes qualitatively similar to the one found in disk (a). On the smooth part of curve (b), flux jumps correspond to a minor fraction of $\phi_0$. Moreover, the first two jumps are negative (see the inset), indicating that the superconducting film *expels* magnetic field when a vortex jumps inside. In principle, the magnetic flux may be expelled if there were vortices of the opposite direction (anti-vortices) frozen inside the film, so that they annihilate with entering vortices. However, no incidentally frozen vortices are present in the case of Figure 1, which was proved by, for example, the symmetric response for the opposite field direction. The influence of the edge[1-4] discussed in the introduction can decrease the amplitude of flux jumps and is partly (see below) responsible for small and non-quantized steps in disk (a). However, the existence of negative flux jumps is unexpected and counter-intuitive.

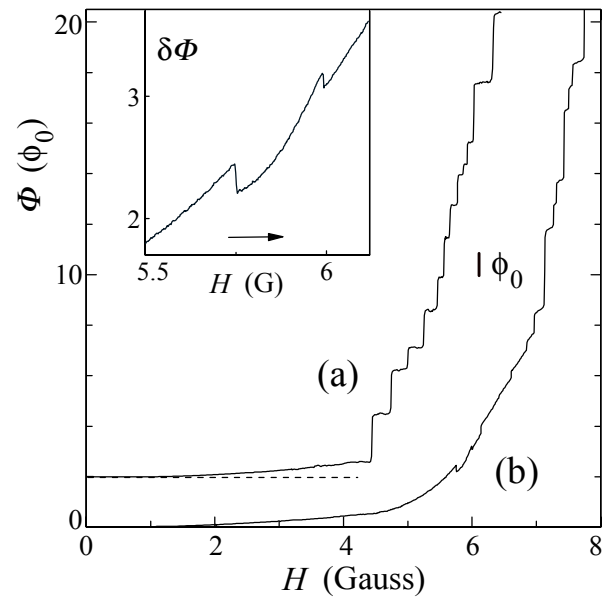

Figure 1. Penetration of perpendicular magnetic field in thin superconducting films. The two curves show the amount of flux $\Phi$ inside as a function of increasing magnetic field $H$ for two aluminium disks of diameter $D \approx 15$ μm and thickness $h \approx 0.1$ μm at temperature $T \approx 0.5$ K. Initially, the samples were cooled in zero field. The measurements have been performed by the ballistic Hall magnetometry (see Fig. 2). For convenience, we define $\Phi$ so that it has zero slope in the low-field limit where $M \propto H$ (such a notion ignores the amount of flux in the λ-layer for the ideal Meissner state; the latter flux is proportional to $H$, and taking it into account would only lead to an additional constant slope for $\Phi$-$H$ curves). The absolute scale of the curves along the $\Phi$-axis is determined with the experimental accuracy of about 10%. Curve (a) is shifted for clarity. The inset magnifies a part of curve (b) exhibiting negative flux jumps.

To understand the origin of the negative flux jumps as well as why similar disks exhibit so different behaviour, we performed a number of dedicated experiments using various sample geometries. The key results are summarised in Figure 2, where we try to maximally simplify the experimental situation by examining only the penetration of the very first vortex and by employing superconducting disks of a relatively small diameter of 2 µm. The advantage of using such small samples is that bulk pinning becomes negligible compared to the interaction of vortices with the edge and, as a result, the first vortex comes right to the disk's centre[8,9]. Therefore, we can study the penetration of an individual vortex at the same, well-defined, distance $D/2$ from the edge. As clearly seen in Fig. 2, the amount of flux carried by entering vortices crucially depends on edge roughness. A superconductor with a smooth edge (as in Fig.2a) exhibits a negative flux jump when the first vortex enters inside, in agreement with the behaviour in Fig.1b. On the other hand, if we introduce a sharp defect at the disk's edge (Fig.2c), the more-or-less expected behaviour for entering vortices is recovered (cf. Fig.1a). Fig. 2b shows an intermediate case of a relatively rough boundary so that an entering vortex brings in no flux. Note that the exiting vortices carry away approximately the same amount of flux ($\approx 0.7\phi_0$), independently of the edge roughness. If no special care is taken[7-9], our microfabricated disks often have a jagged edge or, at least, a few cuts at the edge, which occurs because of ripping of the evaporated film during lift-off. To avoid the ripping, the samples in Fig. 2 were prepared using a thick double-layer resist. A retrospective analysis has shown that disk (a) in Fig 1 has a jagged edge while the edge of disk (b) is smooth.

Figure 2 clearly shows that the negative flux jumps are somehow connected with the stronger bending of the curves in higher fields and the increase of the penetration field $H_p$ for the first vortex. The smoother the edge, the longer the Meissner state (without a vortex) persists and the stronger the Meissner curve bends upwards. The bending means that the magnetic field penetrating in the near-surface layer cannot be described by the London screening model, $B(x) \propto \exp(-x/\lambda)$, which requires a linear dependence $M \propto H$ and, accordingly, $\Phi(H) = 0$ (see caption of Fig. 1). The strong non-linear Meissner effect observed for samples with smooth edges is due to the presence of a high surface barrier which allows a superconductor to persist in a metastable (superheated or supercooled) state[8-10]. Fig. 2 shows that the edge roughness strongly influences the barrier for the vortex entry while this influence on its exit is considerably weaker.

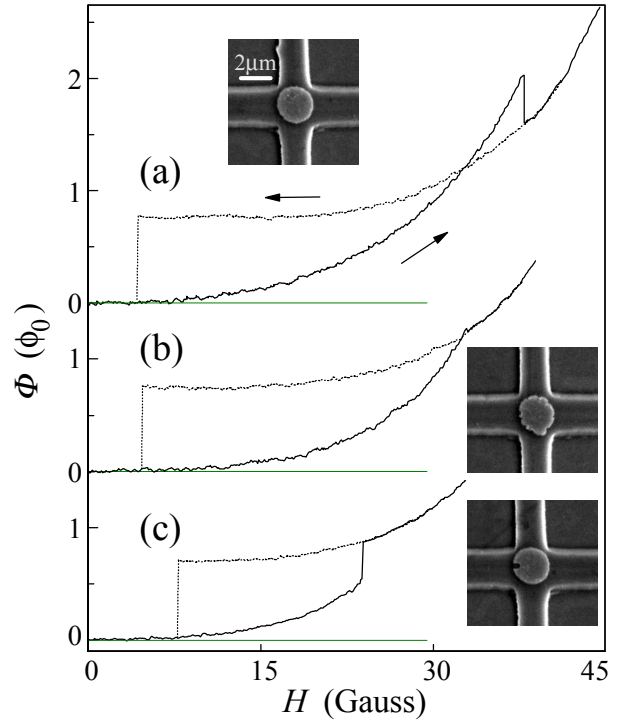

Figure 2. Penetration and expulsion of the very first vortex in superconductors with various edge roughness. Solid and dotted curves are for sweeping the magnetic field up and down, respectively. The hysteresis is due to a surface barrier and, until a vortex enters or leaves the disks, the measured curves are reproducible for sweeps in both directions. Insets show micrographs of the studied Al disks ($D \approx 2$ µm and $h \approx 0.15$ µm; $T \approx 0.5$K). The rough edges for disks (b,c) were intentionally drawn by electron-beam lithography. The disks are placed on top of Hall probes made from a high-mobility two-dimensional electron gas (see the micrographs). Such probes measure the average magnetic field in the central area of the Hall cross[7]. For brevity, one can consider the Hall magnetometers as fluxmeters with a square detection loop, in the center of which a superconducting sample is placed. The superconducting coherence length $\xi(T=0K)$ for the disks' material is $\approx 0.25$ µm and the magnetic penetration length $\lambda(0) \approx 70$ nm, i.e. the material is a type-I superconductor ($\kappa = \lambda/\xi \approx 0.3$). One has to bear in mind that thin films in perpendicular magnetic field behave more like type-II superconductors and exhibit vortex structures[8-11]. We can move into the true type-II regime by using less pure Al and have observed the discussed behaviour also for $\kappa > 1$. However, the reduced screening due to unavoidably larger $\lambda$ for large $\kappa$, led to rapid deterioration of our experimental resolution and, here, we opt to present the data for lower $\kappa$.

In order to explain the origin of the negative flux jumps in superconductors with a smooth edge, it is convenient to refer directly to theoretical curves in Figure 3a. Here we present the magnetization response for a disk with parameters close to those for the disk in Fig. 2a. The curves are found by solving numerically the full 3D set of Ginzburg-Landau equations. In the context of this paper, the important feature to notice is the intersection of magnetization curves for superconducting states with a different number $L$ of vortices. The intersection occurs in a metastable regime and means that, as the field increases, the preceding vortex state can accommodate more flux (i.e. be less diamagnetic) than the following, more energetically favourable state with a larger $L$. This leads to the possibility of vortex jumps that can bring inside any amount of flux from zero to about $\pm\phi_0$, depending on the field $H$ where the vortex entry occurs (see Fig. 3b). The theory allows the persistence of metastable states[10,11] but is not able to predict at which particular field such states become unstable, i.e. at which $H$ a vortex jumps in or out. Among the factors that influence the actual value of $H_p$ are obviously the shape of the edge (Fig. 2) and an unknown mechanism of how a vortex overcomes the surface barrier[11]. In principle, the surface barrier for the penetration of the first vortex exists up to the thermodynamic field $H_C$ [12]. In our experiments, only a part of the full length of theoretical curves is realised. For example, we have never seen the bending and intersection of magnetization curves with decreasing $H$, which could lead to a situation where also exiting vortices carry away a negative flux (see Fig. 3a).

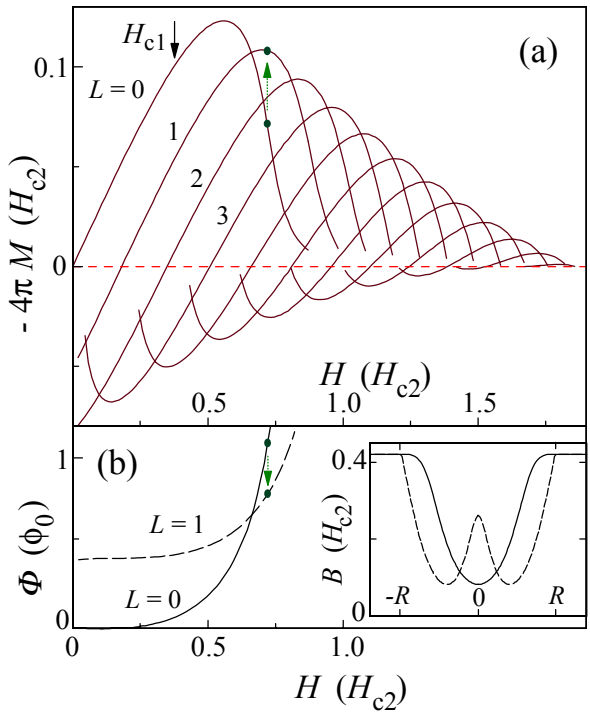

Figure 3. Calculated magnetization response (a) for a disk with $D = 8\xi$, $h = 0.1\xi$ and $\lambda = 0.28\xi$. $L$ is the fluxoid number which, for simplicity[9-11], can be considered as the number of vortices inside the disk. (b) – amount of flux $\Phi$ inside the disk replotted from the $M$-$H$ dependence. The solid arrow marks the low critical field $H_{C1}$ where the Meissner state ($L = 0$) and the state with one vortex ($L=1$) have the same free energy. Markers on the curves with $L = 0$ and 1 illustrate the possibility of the vortex entry which leads to flux expulsion. The inset shows the radial distribution of magnetic field inside a superconducting cylinder ($D = 20\xi$ and $\lambda = 2\xi$) for a similar situation where the entering vortex reduces the amount of flux inside the cylinder. The Meissner state is shown by the solid curve and the $L =1$ state by the dashed one.

To elucidate the physics behind the observed negative flux jumps, it is instructive to consider the distribution of magnetic field before and after such a jump. The inset in Fig. 3b plots the field distribution for the superheated Meissner state and the state with one vortex inside at the same applied field. For clarity, we opt to discuss the case of a superconducting cylinder rather than a disk. The cylindrical geometry keeps the essential physics intact but allows us to avoid a discussion of obscuring demagnetisation effects. It also illustrates the fact that negative flux jumps are a general feature and not exclusive for the thin-film geometry. In the superheated state (see the inset), the external field strongly suppresses the order parameter near the edge, so that the magnetic field penetrates inside the cylinder by as much as a few $\lambda$ deeper than the London model allows. The higher the applied field, the stronger the order parameter is suppressed and the more flux is accumulated in the near-edge layer. On the other hand, when the vortex jumps inside, the screening at the edge notably restores and, accordingly, there is less flux in the near-edge layer. The competition between the flux expelled from the near-edge layer and the flux carried inside by the vortex defines the sign and amplitude of flux jumps. Despite the deceptive simplicity of this explanation, there is no simple way to explain why the vortex entry restores screening while the field at the edge remains the same. It is a non-linear property of the system.

Now we turn to the question why the magnitude of flux jumps is not quantized, even when the surface barrier is suppressed by the edge roughness, and why the distance between the curves with and without a vortex is less than $\phi_0$ (cf. Fig. 2). The latter implies that even for the equilibrium entry and exit, the vortex's flux (and not only the amplitude of the corresponding flux jumps) is considerably less than $\phi_0$. These observations can be explained by changes in the vortex structure due to the finite-size effect[1-4]. For quantitative analysis, Figure 4 plots the amount of flux $\phi$ carried by a vortex versus its distance to the disk edge (i.e. disk radius $D/2$) which is measured in units of the effective penetration length, $\lambda_{eff} = \lambda^2/h$ [3,4]. Because of the surface barrier that is always present even for disks with a rough edge, we cannot directly determine the equilibrium field $H_{C1}$ and, therefore, the amount of flux carried by a vortex in equilibrium. To this end, we notice from our theoretical analysis that $M$-$H$ and $\Phi$-$H$ curves for $L =0$ and 1 are nearly parallel in fields below $H_{C1}$ (see Fig. 3) and, hence, the amount of flux, φ, associated with the vortex exit is sufficiently close – within our experimental uncertainty of 10% - to the flux carried by a vortex in equilibrium. Accordingly, we have measured the amplitude of flux jumps for the vortex exit. All our data for different samples and temperatures fall on a single experimental curve, if plotted against $D/\lambda_{eff}$. Fig. 4 also shows the corresponding dependence found theoretically[13]. There is excellent agreement between the experiment and theory. It is interesting to note that, in a typical experimental situation, superconducting films have $h$ and $\lambda$ of the order of 0.1μm, and it is very unlikely that a vortex can jump farther than 1μm from the edge before being stopped by pinning, even in samples with low pinning. According to Fig. 4, in such a case the flux carried by vortices is reduced to less than half $\phi_0$. Only vortices as far as a hundred micrometers away from the film edge have their flux quantized with accuracy better than 1%.

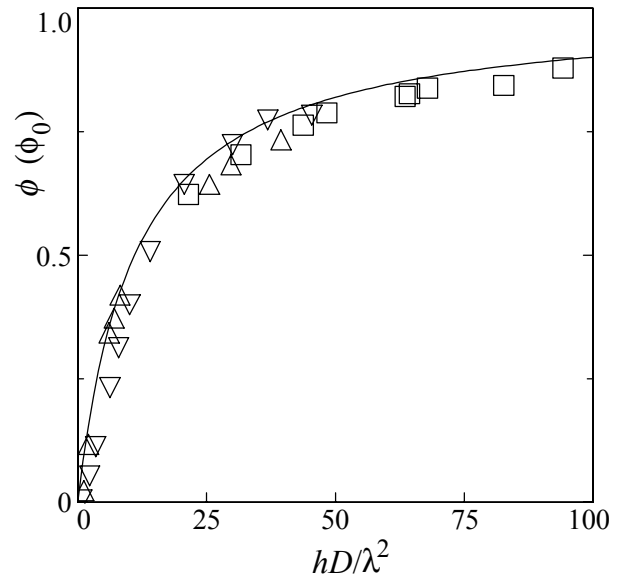

Figure 4. The amount of magnetic flux associated with a vortex in the centre of a superconducting disk in equilibrium. The solid curve is an approximate theoretical dependence $\phi \approx \phi_0\gamma/(\alpha+\gamma)$ found[3,13] numerically in the limit $h \ll \lambda$, where $\gamma=hD/\lambda^2$ and $\alpha \approx 11.0$. Different symbols show experimental data for three disks with $D \approx 2$ ($\triangle$), 2.4 ($\triangledown$) and 4 ($\square$) μm, $h$ from $\approx 0.13$ to $0.17$ μm and the superconducting parameters as in Fig. 2. To obtain different data point for each of the disks, we varied the penetration length $\lambda(T)$ by changing temperature from 0.4 K to close to $T_C \approx 1.25$ K. No fitting parameters were used, except for a slight adjustment ($\leq 10\%$) of the absolute scale along Y-axis for each of the disks.

In summary, two independent reasons give rise to non-quantized penetration of magnetic field in type-II superconductors. The first effect (theoretically established long time ago but never observed and often perceived as small) arises due to changes in the structure of near-edge vortices. This should be important in thin films and, in our opinion, can account for a number of unexplained observations reported in literature. The second, unexpected, effect is more general and appears due to the inevitable presence of surface barriers (for example, the Bean-Livingston barrier). If such barriers are sufficiently high, the induced non-linearity in screening can reach its extreme, causing "negative vortices"; otherwise, the barrier can still deny the quantized penetration. One or both of the above effects can be expected in many – if not the majority of – relevant experimental situations.

We thank V.Kogan for discussions, and FOM, INTAS and NATO for financial support.

geim@sci.kun.nl